%% file: paper_arxiv_second_submission.tex
\documentclass[final]{IEEEtran}
\usepackage{psfrag}
\usepackage{epic,eepic}
\usepackage{graphicx}
\usepackage{psfrag}
\usepackage[dvips]{color}
\usepackage{bm}
\usepackage{amssymb,amsmath}
\usepackage{multirow}
\usepackage{subfigure}
\usepackage{url}
\usepackage[usenames,dvipsnames]{pstricks}
\usepackage{tikz}
\usepackage{pgfplots}
\usepackage[linesnumbered,algoruled]{algorithm2e}

\usepackage[normalem]{ulem}
\usepackage[user]{zref}

\pgfplotscreateplotcyclelist{black white}{%
mark=star\\%
every mark/.append style={fill=gray},mark=*\\%
every mark/.append style={fill=gray},mark=square*\\%
every mark/.append style={fill=gray},mark=triangle*\\%
every mark/.append style={fill=gray},mark=diamond*\\%
every mark/.append style={fill=gray},mark=pentagon*\\%
every mark/.append style={fill=gray},mark=10-pointed star\\%
}

\pgfplotscreateplotcyclelist{mark list}{%
every mark/.append style={solid,fill=.!80!black},mark=*\\%
every mark/.append style={solid,fill=.!80!black},mark=square*\\%
every mark/.append style={solid,fill=.!80!black},mark=triangle*\\%
every mark/.append style={solid,fill=.!80!black},mark=halfsquare*\\%
every mark/.append style={solid,fill=.!80!black},mark=pentagon*\\%
every mark/.append style={solid,fill=.!80!black},mark=halfcircle*\\%
every mark/.append style={solid,fill=.!80!black,rotate=180},mark=halfdiamond*\\%
every mark/.append style={solid,fill=.!80!black!40},mark=otimes*\\%
every mark/.append style={solid,fill=.!80!black},mark=diamond*\\%
every mark/.append style={solid,fill=.!80!black},mark=halfsquare right*\\%
every mark/.append style={solid,fill=.!80!black},mark=halfsquare left*\\%
}

\newcounter{revc}
\makeatletter \zref@newprop{revcontent}{} \zref@addprop{main}{revcontent}
\zref@newprop{revsec}{} \zref@addprop{main}{revsec}

\newcommand{\revi}[2]{%
\zref@setcurrent{revsec}{\thesection}%
\zref@setcurrent{revcontent}{#2}%
\refstepcounter{revc}%
\label{#1}
\zlabel{#1}%
#2
}

\newcommand{\revr}[2]{%
\zref@setcurrent{revsec}{\thesection}%
\zref@setcurrent{revcontent}{#2}%
\refstepcounter{revc}%
\zlabel{#1}%
\label{#1} 
\sout{#2}
} \makeatother

\DeclareMathOperator*{\argmin}{\arg\!\min}
\DeclareMathOperator*{\argmax}{\arg\!\max}

\title{LLR Compression for BICM Systems \\ Using Large Constellations}

\author{\large S. Rosati$^*$, S. Tomasin$^\dag$, M. Butussi$^\ddag$, and B. Rimoldi$^*$\\
\normalsize $^*$\'Ecole Polytechnique F\'ed\'erale de Lausanne, Switzerland\\
$^\dag$University of Padova, Italy\\
$^\ddag$Abilis Systems s.a.r.l., Switzerland\\
Email: \{stefano.rosati@epfl.ch, tomasin@dei.unipd.it, matteo.butussi@abilis.com, bixio.rimoldi@epfl.ch\}
}

\begin{document}

\maketitle

\begin{abstract}
\revi{rev3-1}{
Digital video broadcasting (DVB-C2) and other modern communication standards increase diversity by means of a symbol-level interleaver that spans over several codewords.
De-interleaving at the receiver requires a large memory, which has a significant impact on the implementation cost. In this paper, we propose a technique that reduces the de-interleaver memory
size. 
By quantizing log-likelihood ratios with  bit-specific quantizers and compressing the quantized output, we can significantly reduce the memory size with a negligible increase in computational complexity.
Both the quantizer and compressor are designed via a GMI-based maximization procedure.
For a typical DVB-C2 scenario, numerical results show that the  proposed solution enables a memory saving up to 30\%.}
\end{abstract}

\section{Introduction}
\def\fs{Bit-interleaved coded modulation (BICM) is an effective technique for achieving high communication rates by encoding data bits, interleaving the encoded bits, and then mapping bits into symbols \cite{Zehavi-92,Caire-may98,Guillen-book}. }\fs \revi{rev1-1}{To provide diversity, symbols belonging to different encoded blocks can be interleaved before transmission over correlated fading channels. This is the case, for example, of orthogonal frequency division multiplexing (OFDM) systems when adjacent cells in the frequency domain, or symbols in the time domain, see correlated channels.
In this case it is useful to do symbol-level frequency and time interleaving \cite{DVB-C2, DVB-T2}.} In order to increase the spectral efficiency, large symbol constellations can be used. For example, for the second generation digital video broadcasting standard of cable transmission (DVB-C2) \cite{DVB-C2}, the constellation size is up to 4,096 points and the symbol interleaver is up to 51,776 symbols long; 
its wireless counterpart, DVB-T2 \cite{DVB-T2} uses constellations of a size up to 256 points, with a symbol interleaver that can contain up to 1,023 forward error correction (FEC) codewords; and the Homeplug-AV2 standard \cite{HomeplugAV} for communication over powerline uses a constellation of a size up to 4,096 points.
At the receiver,  a natural choice would be to revert the operation of the transmitter, thus first perform symbol de-interleaving on the demodulated samples, followed by demapping that provides the log-likelihood ratio (LLR) for each encoded bit and then bit de-interleaving before FEC decoding. With long symbol interleavers, these operations require a large amount of memory that has an impact on the cost and on the area of a single-chip receiver. One solution consists of a compact representation of the LLR, which can be obtained by quantization and compression of this information. Note that both the quantization and the compression of an LLR have been investigated  to reduce the memory occupation of systems employing hybrid automatic repeat request (HARQ) \cite{Danieli}, where multiple versions of the same packets must be stored. \revi{rev3-5}{LLR quantization has been investigated for multiple-input multiple-output (MIMO) systems, and a bound on the asymptotic bit error rate (BER) achieved with linear 
binary codes over a flat Rayleigh fading channel has been derived in \cite{Lai-oct12}.} Moreover, LLR compression is used also in compress and forward systems \cite{Haghighat-mar12} and their application to multicell processing \cite{Khattak-2008,Hu-jul06}.

The mutual information (MI) between the transmitted data bits and the compressed words provides a good approximation of what rate can be achieved with practical FEC schemes, and its maximization can be considered as a design criterion for LLR quantization and compression. Since LLRs associated with bits that have been mapped to the same symbol are correlated (as affected by the same noise sample), joint quantization and compression of groups of bits can yield a higher MI.
For example, Danieli et al. proposed applying vector quantization to the LLR \cite{Danieli}, however, this solution becomes infeasible as the constellation size gets larger, and other approaches have been proposed. For a BPSK transmission over the additive white Gaussian Noise (AWGN) channel, the non-uniform LLR quantizer that maximizes the MI is derived by Rave in \cite{Rave-apr09}. By observing that the quantized values are not uniformly distributed, 
Rave suggested applying entropy coding in order to further reduce storage requirements. A suboptimal approach, where MI is maximized under the constraint that all quantized values have the same probability, has been considered in \cite{Novak-ISIT09}, where the analysis is carried out for BPSK transmissions over a Rayleigh fading channel. Indeed, LLR compression is a crucial task in modern communication chips, especially when large blocks of soft bits must be handled, as for low-density parity-check (LDPC) codes.

In this paper, we propose a quantization and compression technique for LLR in systems that uses large constellations and long symbol interleavers. We focus, in particular, on the DVB-C2 system, where the transmitter symbols are interleaved before being mapped on different carriers of multiple OFDM blocks. At the receiver, the samples must be de-interleaved and demapped. In order to reduce the memory occupation, we propose first to demap the received signal and then to perform de-interleaving on groups of LLRs (corresponding to data symbols). In order to ease de-interleaving, the total number of bits representing all the LLRs associated with a single symbol is fixed. In this manner, the symbol de-interleaver moves memory blocks of the same size.
 
To design both quantization and compression,  we use the generalized mutual information (GMI) \cite{Kaplan-93,Martinez-09,Nguyen-feb11}, that provides the achievable throughput, taking into account the approximation occurred in computing the quantized LLRs. Our first contribution is the LLR quantizer design that maximizes the GMI for a given total number of quantization bits among all LLRs. We not only adapt the quantization levels, but also optimize the number of bits used for the representation of the LLR of each bit of the constellation. Our second contribution stems from the observation that quantized LLRs are not uniformly distributed. Therefore, we propose a lossy compression procedure of the quantized LLRs. We begin from an Huffman representation of the quantized LLR. We gather in a word the quantized LLRs associated with a symbol. If this word is longer than a given number of bits, the compressor replaces some quantized values with others that have a shorter representation. We optimize the 
compressor in terms of maximum GMI under the constraint on the total number of bits used to represent a symbol. This is a  multidimensional multiple-choice knapsack (MMCK) problem \cite{Kellerer-book}, for which we derive a suboptimal but practical solution. Finally we present the numerical results for typical DVB-C2 scenarios. 

The rest of the paper is organized as follows. In Section II, we describe the system model and introduce the receiver architecture. In Section III, we provide the details of the design of the quantizer. We describe the lossy compression technique in Section IV. In Section V, we present and discuss numerical results, comparing the various options introduced in the previous section. Lastly we draw some conclusions in Section VI.

\section{System Model}

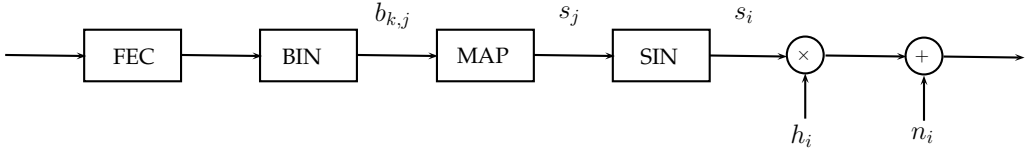
\begin{figure*}
\centering
{
\scalebox{0.70}{
\begin{pspicture}(0,-1.3592187)(19.38,1.3592187)
\psframe[linewidth=0.04,dimen=outer](3.36,0.90578127)(1.48,-0.11421875)
\psframe[linewidth=0.04,dimen=outer](6.7,0.90578127)(4.82,-0.11421875)
\psline[linewidth=0.04cm,arrowsize=0.05291667cm 2.0,arrowlength=1.4,arrowinset=0.4]{->}(3.34,0.40578124)(4.88,0.38578126)
\psframe[linewidth=0.04,dimen=outer](10.06,0.90578127)(8.18,-0.11421875)
\psframe[linewidth=0.04,dimen=outer](13.4,0.90578127)(11.52,-0.11421875)
\psline[linewidth=0.04cm,arrowsize=0.05291667cm 2.0,arrowlength=1.4,arrowinset=0.4]{->}(10.04,0.40578124)(11.58,0.38578126)
\psline[linewidth=0.04cm,arrowsize=0.05291667cm 2.0,arrowlength=1.4,arrowinset=0.4]{->}(13.38,0.39578125)(14.85,0.37578124)
\psline[linewidth=0.04cm,arrowsize=0.05291667cm 2.0,arrowlength=1.4,arrowinset=0.4]{->}(6.68,0.40578124)(8.22,0.38578126)
\psline[linewidth=0.04cm,arrowsize=0.05291667cm 2.0,arrowlength=1.4,arrowinset=0.4]{->}(0.0,0.40578124)(1.54,0.38578126)
\psline[linewidth=0.04cm,arrowsize=0.05291667cm 2.0,arrowlength=1.4,arrowinset=0.4]{->}(15.58,0.39578125)(17.12,0.37578124)
\usefont{T1}{ppl}{m}{n}
\rput(2.4482813,0.3746875){\large FEC}
\usefont{T1}{ppl}{m}{n}
\rput(5.6340623,0.37453124){\large BIN}
\usefont{T1}{ppl}{m}{n}
\rput(9.072812,0.39546874){\large MAP}
\usefont{T1}{ppl}{m}{n}
\rput(12.422031,0.37453124){\large SIN}
\usefont{T1}{ppl}{m}{n}
\rput(7.364531,1.1557813){\Large $b_{k,j}$}
\usefont{T1}{ppl}{m}{n}
\rput(14.034532,1.1557813){\Large$s_i$}
\usefont{T1}{ppl}{m}{n}
\rput(10.714531,1.1557813){\Large $s_j$}
\pscircle[linewidth=0.04,dimen=outer](15.194531,0.40578124){0.37}
\usefont{T1}{ptm}{m}{n}
\rput(15.164532,0.40546876){$\times$}
\psline[linewidth=0.04cm,arrowsize=0.05291667cm 2.0,arrowlength=1.4,arrowinset=0.4]{->}(15.194531,-0.80421877)(15.194531,0.05578125)
\pscircle[linewidth=0.04,dimen=outer](17.45,0.38578126){0.37}
\usefont{T1}{ptm}{m}{n}
\rput(17.411875,0.38296875){$+$}
\psline[linewidth=0.04cm,arrowsize=0.05291667cm 2.0,arrowlength=1.4,arrowinset=0.4]{->}(17.45,-0.84421873)(17.45,0.01578125)
\psline[linewidth=0.04cm,arrowsize=0.05291667cm 2.0,arrowlength=1.4,arrowinset=0.4]{->}(17.82,0.37578124)(19.36,0.35578126)
\usefont{T1}{ppl}{m}{n}
\rput(15.124531,-1.1242187){\Large $h_i$}
\usefont{T1}{ppl}{m}{n}
\rput(17.424532,-1.1242187){\Large $n_i$}
\end{pspicture} 
}
}
\caption{Transmitter and channel models. FEC: Forward Error Correction; BIN: Bit Interleaver; MAP: Mapper; SIN: Symbol Interleaver.}
\label{fig1}
\end{figure*}

We consider the transmission scheme of Fig. \ref{fig1}, where data bits are encoded by FEC. Bit-interleaving (BIN) and Mapping (MAP) of bits to $M$-QAM symbols follow. Encoded bits are indicated as $b_{k,j} \in \{0,1\}$, where\footnote{In this paper $\log x$ denotes the base-2 logarithm of $x$, and $\ln x$ denotes the base-$e$ logarithm.} $k=1, 2, \ldots, \log M$, and $j$ is the index of the QAM symbol $s_j$. \revi{rev1-1.2}{The generated symbols are then interleaved (SIN) before transmission to provide diversity over correlated fading channels.} \revi{rev3-7}{The symbol that has position $j$ within the block entering the symbol interleaver is moved to position $i = \mathcal M(j)$, where  $\mathcal M$ is a permutation over the index set.}

The symbol $s_i$ is transmitted on a fading channel, i.e., it is multiplied by the channel gain $h_i$. Then complex white Gaussian noise (AWGN) $n_i$ is added. The noise has zero mean and power $\sigma^2$. With this model, we appropriately describe the main features of many communication systems, including those based on OFDM\footnote{If the cyclic prefix is longer than the channel impulse response and assuming perfect synchronization, the cascade of OFDM modulation, the  channel, and  OFDM demodulation is equivalent to a set of parallel memoryless fading channels, each with a different gain $h_i$.}. Single carrier transmissions with linear equalization, as well as MIMO systems with linear receivers can be cast into this model. Hereafter, we assume that the channel gains $h_i$ are known to the receiver.

\subsection{Receiver Implementation}

 \begin{figure*}
 \centering
\subfigure[Conventional Receiver.]{
\scalebox{0.65}{
\begin{pspicture}(0,-1.8367188)(18.67,1.8767188)
\psframe[linewidth=0.04,dimen=outer](3.77,1.3932812)(1.89,0.37328124)
\psframe[linewidth=0.04,dimen=outer](7.11,1.3932812)(5.23,0.37328124)
\psline[linewidth=0.04cm,arrowsize=0.05291667cm 2.0,arrowlength=1.4,arrowinset=0.4]{->}(3.75,0.8932812)(5.29,0.87328124)
\psframe[linewidth=0.04,dimen=outer](10.47,1.3932812)(8.59,0.37328124)
\psframe[linewidth=0.04,dimen=outer](13.81,1.3932812)(11.93,0.37328124)
\psline[linewidth=0.04cm,arrowsize=0.05291667cm 2.0,arrowlength=1.4,arrowinset=0.4]{->}(10.45,0.8932812)(11.99,0.87328124)
\psframe[linewidth=0.04,dimen=outer](17.15,1.3932812)(15.27,0.37328124)
\psline[linewidth=0.04cm,arrowsize=0.05291667cm 2.0,arrowlength=1.4,arrowinset=0.4]{->}(13.79,0.8932812)(15.33,0.87328124)
\psline[linewidth=0.04cm,arrowsize=0.05291667cm 2.0,arrowlength=1.4,arrowinset=0.4]{->}(7.09,0.8932812)(8.63,0.87328124)
\psline[linewidth=0.04cm,arrowsize=0.05291667cm 2.0,arrowlength=1.4,arrowinset=0.4]{->}(0.41,0.8932812)(1.95,0.87328124)
\psline[linewidth=0.04cm,arrowsize=0.05291667cm 2.0,arrowlength=1.4,arrowinset=0.4]{->}(17.11,0.8932812)(18.65,0.87328124)
\psframe[linewidth=0.04,linestyle=dashed,dash=0.16cm 0.16cm,dimen=outer](5.264531,-0.5367187)(3.8245313,-1.8367188)
\psline[linewidth=0.04cm,linestyle=dashed,dash=0.16cm 0.16cm,arrowsize=0.05291667cm 2.0,arrowlength=1.4,arrowinset=0.4]{<->}(4.5245314,-0.57671875)(4.5645313,0.86328125)
\usefont{T1}{ppl}{m}{n}
\rput(4.514531,-1.1267188){$M_{\mathrm{SD}}^{a}$}
\psframe[linewidth=0.04,linestyle=dashed,dash=0.16cm 0.16cm,dimen=outer](12.224531,-0.51671875)(10.784532,-1.8167187)
\psline[linewidth=0.04cm,linestyle=dashed,dash=0.16cm 0.16cm,arrowsize=0.05291667cm 2.0,arrowlength=1.4,arrowinset=0.4]{<->}(11.484531,-0.55671877)(11.524531,0.88328123)
\usefont{T1}{ppl}{m}{n}
\rput(2.9415624,0.8621875){QUA}
\usefont{T1}{ppl}{m}{n}
\rput(6.0382814,0.8620312){SDI}
\usefont{T1}{ppl}{m}{n}
\rput(9.488125,0.8829687){DEM}
\usefont{T1}{ppl}{m}{n}
\rput(12.834687,0.8620312){BDI}
\usefont{T1}{ppl}{m}{n}
\rput(16.174688,0.93328124){DEC}
\usefont{T1}{ppl}{m}{n}
\rput(11.474531,-1.0867188){$M_{\mathrm{BD}}$}
\usefont{T1}{ppl}{m}{n}
\rput(0.81453127,1.6732812){$ r_i, h_i$}
\usefont{T1}{ppl}{m}{n}
\rput(11.264531,1.6732812){$\lambda_{k,j}$}
\usefont{T1}{ppl}{m}{n}
\rput(7.824531,1.6732812){$ r_j, H_j$}
\end{pspicture} 
}}
\subfigure[Proposed Receiver.]{
\scalebox{0.6}{
\begin{pspicture}(0,-1.8192188)(28.38547,1.8592188)
\psframe[linewidth=0.04,dimen=outer](6.8254685,1.3807813)(4.945469,0.36078125)
\psframe[linewidth=0.04,dimen=outer](10.165469,1.3807813)(8.285469,0.36078125)
\psline[linewidth=0.04cm,arrowsize=0.05291667cm 2.0,arrowlength=1.4,arrowinset=0.4]{->}(6.8054686,0.8807813)(8.3454685,0.86078125)
\psframe[linewidth=0.04,dimen=outer](13.525469,1.3807813)(11.645469,0.36078125)
\psframe[linewidth=0.04,dimen=outer](16.865469,1.3807813)(14.985469,0.36078125)
\psline[linewidth=0.04cm,arrowsize=0.05291667cm 2.0,arrowlength=1.4,arrowinset=0.4]{->}(13.505468,0.8807813)(15.045468,0.86078125)
\psframe[linewidth=0.04,dimen=outer](20.20547,1.3807813)(18.325468,0.36078125)
\psline[linewidth=0.04cm,arrowsize=0.05291667cm 2.0,arrowlength=1.4,arrowinset=0.4]{->}(16.845469,0.8807813)(18.38547,0.86078125)
\psline[linewidth=0.04cm,arrowsize=0.05291667cm 2.0,arrowlength=1.4,arrowinset=0.4]{->}(10.145469,0.8807813)(11.685469,0.86078125)
\psline[linewidth=0.04cm,arrowsize=0.05291667cm 2.0,arrowlength=1.4,arrowinset=0.4]{->}(3.4654686,0.8807813)(5.005469,0.86078125)
\psline[linewidth=0.04cm,arrowsize=0.05291667cm 2.0,arrowlength=1.4,arrowinset=0.4]{->}(20.165468,0.8807813)(21.70547,0.86078125)
\psframe[linewidth=0.04,linestyle=dashed,dash=0.16cm 0.16cm,dimen=outer](21.720001,-0.51921874)(20.28,-1.8192188)
\psline[linewidth=0.04cm,linestyle=dashed,dash=0.16cm 0.16cm,arrowsize=0.05291667cm 2.0,arrowlength=1.4,arrowinset=0.4]{<->}(20.980001,-0.5592188)(21.019999,0.8807812)
\usefont{T1}{ppl}{m}{n}
\rput(20.944532,-1.0692188){$M_{\mathrm{BD}}$}
\psframe[linewidth=0.04,linestyle=dashed,dash=0.16cm 0.16cm,dimen=outer](14.95,-0.47921875)(13.510001,-1.7792188)
\psline[linewidth=0.04cm,linestyle=dashed,dash=0.16cm 0.16cm,arrowsize=0.05291667cm 2.0,arrowlength=1.4,arrowinset=0.4]{<->}(14.21,-0.51921874)(14.25,0.92078125)
\usefont{T1}{ppl}{m}{n}
\rput(14.174531,-1.0692188){$M_{\mathrm{SD}}^{b}$}
\usefont{T1}{ppl}{m}{n}
\rput(15.882656,0.8596875){SDI}
\usefont{T1}{ppl}{m}{n}
\rput(5.8701563,0.8595313){DEM}
\usefont{T1}{ppl}{m}{n}
\rput(9.216719,0.8704688){QUA}
\usefont{T1}{ppl}{m}{n}
\rput(19.235937,0.8596875){UCOM}
\usefont{T1}{ppl}{m}{n}
\rput(0.81453127,1.6557813){$ r_i, h_i$}
\usefont{T1}{ppl}{m}{n}
\rput(7.474531,1.6557813){$\lambda_{k,i}$}
\usefont{T1}{ppl}{m}{n}
\rput(10.944531,1.6557813){$v_{k,i}$}
\psframe[linewidth=0.04,dimen=outer](23.525469,1.3807813)(21.64547,0.36078125)
\psframe[linewidth=0.04,dimen=outer](26.865469,1.3807813)(24.985468,0.36078125)
\psline[linewidth=0.04cm,arrowsize=0.05291667cm 2.0,arrowlength=1.4,arrowinset=0.4]{->}(23.505468,0.8807813)(25.04547,0.86078125)
\psline[linewidth=0.04cm,arrowsize=0.05291667cm 2.0,arrowlength=1.4,arrowinset=0.4]{->}(26.825468,0.8807813)(28.365469,0.86078125)
\usefont{T1}{ppl}{m}{n}
\rput(22.507969,0.8595313){BDI}
\usefont{T1}{ppl}{m}{n}
\rput(25.858438,0.8596875){DEC}
\usefont{T1}{ppl}{m}{n}
\rput(12.466875,0.8596875){COM}
\usefont{T1}{ppl}{m}{n}
\rput(14.224531,1.6557813){$\hat{v}_{k,i}$}
\usefont{T1}{ppl}{m}{n}
\rput(17.684532,1.6557813){$\hat{v}_{k,j}$}
\usefont{T1}{ppl}{m}{n}
\rput(20.98453,1.6557813){$\hat{v}_{k,j}$}
\psframe[linewidth=0.04,dimen=outer](3.5054688,1.3807813)(1.6254687,0.36078125)
\psline[linewidth=0.04cm,arrowsize=0.05291667cm 2.0,arrowlength=1.4,arrowinset=0.4]{->}(0.14546876,0.8807813)(1.6854688,0.86078125)
\usefont{T1}{ppl}{m}{n}
\rput(2.5567186,0.8604688){QUA}
\end{pspicture} } }
\caption{Receiver architectures. QUA: Quantizer; SDI: Symbol De-interleaver; DEM: Demodulator; BDI: Bit De-interleaver; DEC: Decoder; COM: Compression; UCOM: Uncompression.}
\label{fig2}
\end{figure*}
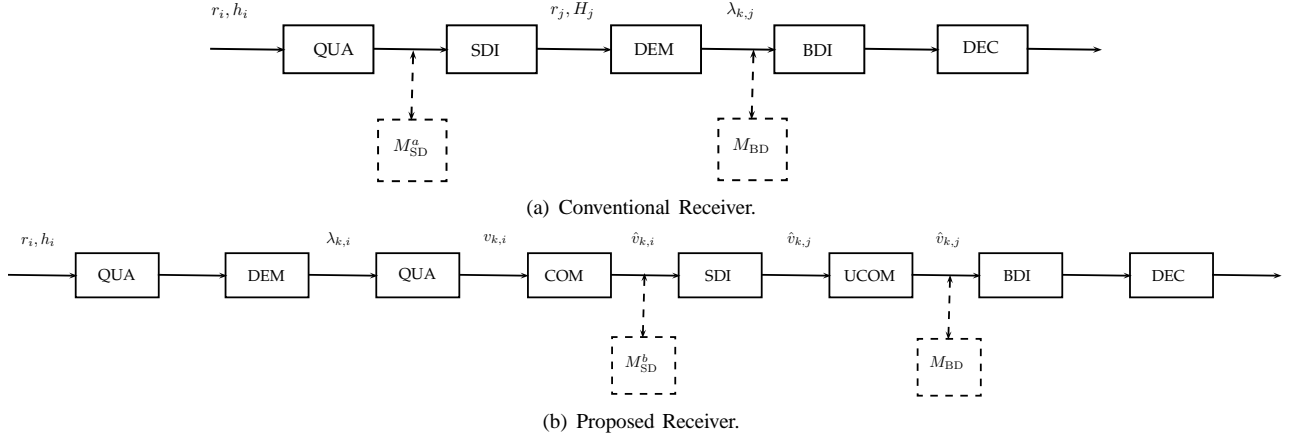

We consider the two receiver alternatives depicted in Fig. \ref{fig2}.
  
\paragraph*{Conventional Receiver} In this receiver  -- depicted in Fig. \ref{fig2} (a) -- the received samples $r_i$ and the channel gains $h_i$ are first de-interleaved (SDI) and then passed to the demapper (DEM) to obtain the LLR $\lambda_{k,j}$ associated with the encoded bit $b_{k,j}$. For an implementation of the receiver on a chip, the received samples, channel gains and LLRs will be represented as quantized values; in particular, quantization is explicitly shown in the figure by block QUA. The quantized LLRs are passed to a bit de-interleaver (BDI) and then to the FEC decoder (DEC) for error correction. In this implementation two blocks of memory, named $M_{\mathrm{SD}}^{a}$ and $M_{\mathrm{BD}}$, are needed. $M_{\mathrm{SD}}^{a}$ is associated with SDI and stores both the received samples and channel gains. $M_{\mathrm{BD}}$, which is associated with BDI, stores LLRs.

\paragraph*{Proposed Receiver} 
In this receiver  -- depicted in Fig. \ref{fig2} (b) -- in order to reduce both the complexity of the de-interleaver and the  total memory, demapping and symbol de-interleaving are swapped. 
\revi{revE3}{The sample $r_i$ is first quantized and then demapped to obtain LLRs $\lambda_{k,i}$, associated with the encoded bit $b_{k,i}$, for $k=1, 2, \ldots, \log M$.  LLR $\lambda_{k,i}$ is further quantized into one of the $L_k$ possible quantization levels, and then the index $v_{k,j}$ of the quantized level associated with the LLR is stored. We assume that the first quantization on the received sample is very precise so that this quantization error can be ignored in the system analysis. This is a reasonable assumption because the quantized received samples are not stored, hence a fine quantization has no drawback on the memory size.} In this implementation the symbol de-interleaver operates on {\em words} of quantized LLRs instead of the quantized received samples. Each word consists of LLRs of bits mapped to the same symbol. 
We compress the quantized LLR values, thus obtaining a smaller memory and at the same time reducing memory swapping operations for the de-interleaver. 
In particular, we consider two components that perform compression (COM) and decompression (UCOM) of the quantized levels $v_{k,j}$.  We observe that a simple implementation of the de-interleaver requires that all compressed words are represented by the same number of bits $\bar{N}$. In this case, symbol de-interleaving boils down to the permutation of blocks of memory of the same size. In order to ensure that compression generates at most $\bar{N}$ bits for each transmitted symbol, we allow for losses, i.e., quantized indices $v_{k,j}$ could be substituted by other indices $\hat{v}_{k,j}$ represented by fewer bits. After symbol de-interleaving, LLR words are uncompressed into fixed-length quantization indices $v_{k,i}$ to allow for bit-de-interleaving (BDI), and they are finally mapped into quantized LLR values $\hat{\lambda}_{k,i}$ before being passed to the FEC decoder.
Also in this case we need two blocks of memory, $M_{\mathrm{SD}}^{b}$, and $M_{\mathrm{BD}}$, both storing LLR quantized LLR levels.
\revi{revE1}{Note that interleavers are often designed in order to operate in a pipelined fashion without the need to double the memory size. 
This is a common feature in today's communication systems, as is the case of DVB-C2  \cite{DVB-C2_guidelines}. 
Combining SDI and BDI in a single de-interleaver would break this feature and then would require a larger memory, thus it is convenient to keep SDI and BDI as two separate operations, even when compression is not employed.}

\revi{rev1-2}{The two architectures of Fig. \ref{fig2} can be compared in two respects: from a complexity and from a memory point of view.
The proposed implementation requires additional complexity for compression/decompression, and less memory for de-interleaving. This complexity  increase can be kept negligible with respect to the decoding complexity. So it has a negligible effect on the receiver cost. 
But, the de-interleaving part has a significant impact on the total memory size, due to the large size of the interleavers used in DVB-C2 \cite{DVB-C2}. 
As shown in Section \ref{memcomp}, the proposed scheme results in a memory reduction of about 30\% with respect to that of the conventional receiver.}

\subsection{LLR Statistics}

Assume equal probability for all constellation points. The minimum distance approximation of the LLR is given by \cite{Viterbi-feb98}
\begin{equation}\label{Approx}
\begin{split}
\lambda_{k,i} = &  - \frac{1}{\sigma^2} \left( \min_{s \in  \mathcal{S}_k(1)}
\left\{ ||r_i - h_i s||^2\right\}  - \right.\\ 
& \left.\min_{s \in \mathcal{S}_k(0)} \left\{||r_i - h_i s||^2\right\} \right)\,,
\end{split}
\end{equation}
where $\mathcal{S}_k(u)$ is the set of constellation points with the $k$-th bit equal to $u \in \{0,1\}$. 
As both LLR computation and compression operate at a symbol level,  unless explicitly required in the following, we drop the symbol index  $i$ or $j$. In this paper, we consider the minimum distance approximation of the LLR, although the proposed solution applies also to other approximations of the LLR (including its exact definition).
As the same value of $\lambda_k$ can be achieved with various values of $r$, the LLRs conditioned on the $b_k$ and $h$ are distributed as a piecewise Gaussian mixture \cite{Alvarado-09}.
The real axis is partitioned into $U$ intervals $[h a_u^{L} ,  h a_u^{U}]$ for $u=1, 2, \ldots, U$. We have
\begin{equation}
\begin{split}
p_{\Lambda_k|B_k,H}(\lambda_k|b_k,h) =  &\sum_{\mu =1}^{G_u} \frac{1}{G_u} \frac{\sigma}{\sqrt{2 \pi}  h \gamma_{\mu, u,k }} \times \\
&  \exp\left[-\frac{1}{2}\left(\frac{\lambda_k \sigma^2- h^2 m_{\mu,u,k}}{ \sigma h \gamma_{\mu, u,k }}\right)^2 \right]\,, 
\\ & \lambda_k \in [h a_u^{L} , h a_u^{U}] \,,
 \end{split}
\label{PDFawgn}
\end{equation}
where $\Lambda_k$, $B_k$, and, $H$ are the random variables corresponding to LLRs, bits, and channel gains respectively, and with $\lambda_k$, $b_k$, and $h$ we denote their realizations.
Note that $\gamma_{1, u,k },\ldots, \gamma_{G_u, u,k }$, and $m_{1,u,k},\ldots, m_{G_u,u,k}$, are the Gaussian mixture parameters of the $u$-th interval, which are also functions of $b_k$. 
In \cite{Alvarado-09} explicit expressions of $p_{\Lambda_k|B_k}(\lambda_k|b_k)$ are derived for squared QAM constellations. 
In the following, we will also need $p_{\Lambda_k|B_k,H}(\lambda_k|b_k,h)$, that can be  obtained by averaging (\ref{PDFawgn}) over the channel PDF, i.e.,
\begin{equation}
p_{\Lambda_k|B_k}(\lambda_k|b_k) = {\rm E}[p_{\Lambda_k|B_k,H}(\lambda_k|b_k,H)] \,,
\label{PDFfading}
\end{equation}
where ${\rm E}[\cdot]$ denotes expected value.%
\revi{revE-m1}{In this paper, we consider two channels: the AWGN channel, where $p_{\Lambda_k|B_k}(\lambda_k|b_k) = p_{\Lambda_k|B_k,H}(\lambda_k|b_k,1)$, and the Rayleigh fading channel, for which a closed form expression of (\ref{PDFfading}) has been derived in \cite{Szczecinski-09}.}\footnote{Note that by assuming large symbol interleaving, uncorrelated channel realizations of symbols belonging to FEC block can be achieved.}

\section{LLR Quantization}

The LLRs associated with the same transmitted symbol are correlated random variables as they are affected by the same noise sample. Therefore, vector quantization \cite{Gersho-book} of the LLR vector $\lambda_1, \lambda_2, \ldots, \lambda_{\log M}$, can be applied. However, this technique is exceedingly complex for large $M$.

Here, we propose instead that the LLR of each bit is quantized by a tailored  quantizer. In fact, each of the $\log M$ LLRs has a different statistic, as shown in (\ref{PDFawgn}), and a great performance benefit can be  achieved by considering $\log M$ quantizers, each with its own quantization intervals.  As noted above, the statistics of the LLR depend on the channel $h	$, therefore adapting the quantizer to the channel associated with the LLR would also increase the accuracy of the quantized representation. However, the decoder should then know also the channel gain, and additional memory should be reserved to store this information. In order to reduce memory occupation, we consider here a scenario where channel gains are discarded after the  LLR computation, and the quantizers are not adapted to the channel levels.

\revi{revE2}{Note that we store the indices that describe the quantized values. The decoder uses a look-up table to determine the quantized values and performs arithmetic operations on these quantities, represented by fixed-point numbers with the same precision for all LLRs.
We assume this  fixed-point representation to be sufficiently accurate to have a negligible effect on the performance. This is realistic because the number of bits used for this representation (internal to the decoder) does not affect the overall memory size.
Thus, we ignore this step and, in the following, focus on quantization only.}

\subsection{Quantization Procedure}

We focus on the uniform quantization of the LLRs, although the derivations are easily extended  to  non-uniform quantization. In particular, the LLR of the $k$-th bit is quantized by a uniform quantizer, having quantization step $q_k$ and $L_k = 2^{w_k}$ levels, where $w_k$ is the number of bits used to describe a level.  The $L_k$ quantization intervals are 
\begin{equation}
\mathcal D_{\ell} = [d_{k, \ell-1}, d_{k, \ell})\,,
\end{equation}
with $\ell=1, 2, \ldots, L_k$, where $d_{k, 0} = -\infty$,  $d_{k, L_k} = \infty$, and
\begin{equation}
d_{k, \ell} = \left(\ell - \frac{L_k}{2}\right) q_k  \,,\quad  \ell=1,2,  \ldots, L_k-1\,.
\end{equation}
Note that $w_k$ and $q_k$ fully specify the quantizer for $\lambda_k$. The quantization process is described as follows:
\begin{equation}
\lambda_k \mbox{ is mapped to index $v_k=\ell$ if } \lambda_k  \in \mathcal D_{\ell} \,.
\label{quant}
\end{equation} 
For each index $v_k$, we have a corresponding quantized LLR value $\lambda^{\rm (Q)}_{v_k, k}$. 
Let the discrete random variable  $V_k$ be the quantization level index of $\Lambda_k$ and  $p_{V_k|B_k}(v_k | b_k)$ the conditional probability mass function (PMF) of  $V_k$, given $B_k$, which can be written as 
\begin{equation}
p_{V_k|B_k}(v_k | b_k) =  \int_{d_{k, v_k-1}}^{d_{k, v_k}} p_{\Lambda_k|B_k}(\lambda_k|b_k)\, {\rm d}\lambda_k\,.
\label{pikgb}
\end{equation}
The unconditional PMF is given by 
\begin{equation} \label{pvk}
p_{V_k}(v_k) = {\rm E}\left[p_{V_k|B_k}(v_k|B_k)\right].
\end{equation}

In general, numerical methods must be used to compute (\ref{pikgb}). For AWGN channels with a fixed channel gain $h$ and given noise power $\sigma^2$, from (\ref{PDFawgn}) we have a closed form expression of the conditional PMF, i.e.,
\begin{equation} \label{pvkbk}
\begin{split} 
p_{V_k|B_k}(v_k | b_k) =  & \sum_{u=1}^U \sum_{\mu =1}^{G_u} \frac{1}{G_u} \left( {\rm Q}\left(\frac{\alpha_u \sigma^2 - m_{\mu, u, k}h^2}{\gamma_{\mu, u, k}\sigma h} \right)  \right. \\
& \left. -   {\rm Q}\left(\frac{\beta_u \sigma^2- m_{\mu, u, k}h^2}{\gamma_{\mu, u, k}\sigma h} \right)  \right)\,,
\end{split}
\end{equation}
where ${\rm Q}(\cdot)$ is the Q-function and
\begin{equation}
\alpha_u = \min \{ \max \{ d_{k, v_k}, h a_u^{L} \} , h a_u^{U} \} 
\end{equation}
\begin{equation}
\beta_u = \min\{ \max\{d_{k, v_k-1}, h a_u^{L}\}, h a_u^{U}\}\,.
\end{equation}

\subsection{Quantization Design}

As performance measure for the design of the quantizer we consider the GMI, defined, for a specific decoder metric, as the supremum among all rates for which the random coding exponent is strictly positive.

In \cite{Jaldeen-10} it has been proved that  the GMI can be upper bounded by
\begin{equation}\label{eq:GMI}
{\rm GMI} \leq \max_{x>0} \sum_{k=1}^{\log M} {\rm BGMI}_k(x)\,,
\end{equation}
where the binary GMI (BGMI) is
\begin{equation}
\begin{split}
{\rm BGMI}_k(x) = &  - {\rm E} \left[ \sum_{b=0}^1 p_{B_k}(b) \times \right.  \\ 
& \left. \log \left( p_{B_k}(b)  + p_{B_k}(1-b) e^{-(2b-1)\hat{\Lambda}_k x} \right) \right]\,.
 \label{BGMIdef}
\end{split}
\end{equation}
Also, note that a suitable mapping can be applied on $\hat{\lambda}_k$ such that (\ref{eq:GMI}) holds with equality, which also occurs when exact LLR is used instead of the approximated LLR. In this case, any rate below the GMI is achievable without the ideal interleaver assumption \cite{Martinez-09}.
Considering the quantization rule in (\ref{quant}) and equiprobable bits, we can rewrite (\ref{BGMIdef}) as
\begin{equation}
\begin{split}
{\rm BGMI}_k(x) = & 1 - \sum_{v_k=1}^{L_k} \frac{1}{2} \left[ p_{V_k|B_k}(v_k|0) \log \left(1+ e^{ \lambda^{\rm (Q)}_{v_k, k} x} \right) +\right. \\ 
& \left. + p_{V_k|B_k}(v_k|1) \log \left(1+ e^{-\lambda^{\rm (Q)}_{v_k, k} x} \right)\right]\,.
\end{split}\label{BGMIdef2}
\end{equation}
We first note that the quantized LLR value that maximizes the GMI can be obtained by setting to zero the derivative of the BGMI with respect to $\hat{\lambda}_k(v_k)$. Doing so yields
\begin{equation}
\lambda^{\rm (Q)}_{v_k, k} = \frac{1}{x} \ln \left(\frac{p_{V_k|B_k}(v_k|1)}{p_{V_k|B_k}(v_k|0)} \right)\,.
\label{optq}
\end{equation}
Inserting \eqref{optq} into \eqref{BGMIdef} we obtain
\begin{equation}
\begin{split}
{\rm BGMI}_k & =  I(B_k; V_k) \\ & =  \frac{1}{2} \sum_{v_k=1}^{L_k} \sum_{b_k=0}^1 p_{V_k|B_k}(v_k | b_k) \log \frac{p_{V_k|B_k}(v_k | b_k)}{p_{V_k}(v_k)}\,,
 \end{split}
\label{max1}
\end{equation}
which coincides with the MI between $b_k$ and $v_k$, and does not depend on $x$.
Substituting \eqref{max1} in \eqref{eq:GMI} yields
\begin{equation}
\begin{split}
{\rm GMI} = & \sum_{k=1}^{\log M} I(B_k; V_k)  \\ 
 = & \frac{1}{2} \sum_{k=1}^{\log M} \sum_{v_k=1}^{L_k} \sum_{b_k=0}^1 p_{V_k|B_k}(v_k | b_k) \log \frac{p_{V_k|B_k}(v_k | b_k)}{p_{V_k}(v_k)} \,.
\label{max2}
\end{split}
\end{equation}
Hence, the GMI is given by the sum over $k$ of the MI between $B_k$ and $V_k$. 

\revi{rev3-17}{Note that in the literature the GMI is proposed as an accurate performance measure for BICM systems with mismatched decoders \cite{Kaplan-93,Martinez-09,Nguyen-feb11}. In the context of this paper, when quantization is performed, the decoder is  mismatched for two reasons: 
(i) the  LLRs coming from a given symbol are not independent as inherently assumed by the decoder; and (ii) the decoder assumes unquantized LLRs.}

\revi{rev2-3}{In order to maximize the GMI, for the design of both the quantizer and the compressor, we choose the quantized value according to (\ref{optq}) with $x =1$ (from the above discussion, any value $x \neq 0$ yields the same GMI).}
As mentioned, we are using a specific quantizer for each of the $\log M$ bits mapped to a symbol. Therefore the objective of the quantization design is to optimize both vectors $\bm{w} = (w_1, w_2, \ldots, w_{\log M})$ and $\bm{q} = (q_1, q_2, \ldots, q_{\log M})$, where $w_k$ and $q_k$ are the bit-length and quantization steps of the quantizer that operates on the LLR of the $k$-th bit.
The quantizer design aims at maximizing the GMI in (\ref{max2}) with the constraint of using $W$ bits for the quantization of all LLRs of a word. Mathematically, we aim at solving
\begin{subequations}
\begin{equation}
\max_{  \substack{\bm{q} , \bm{w}}} \sum_{k=1}^{\log M} I(B_k; V_k)\,,
\label{Lag1}
\end{equation}
s.t.
\begin{equation}
\sum_{k=1}^{\log M} w_k = W\,.
\label{totbits}
\end{equation}
\label{probmax}
\end{subequations}

Unfortunately, the constrained maximization (\ref{probmax}) is a mixed integer programming (MIP) problem and cannot be solved in closed form. We must resort to numerical methods to optimize both $\bm{q}$ and $\bm{w}$.

\subsubsection*{Optimization of the quantization steps $q_k$}

For each  $k=1, 2, \ldots, \log M$, and $w_k=1, 2, \ldots, W$,  we first find the best $q_k$ that maximizes the BGMI, i.e.,
\begin{equation}\label{eq:Sk}
 q_k(w_k)= \argmax_{q_k} I(B_k; V_k) \,.
\end{equation}
The above optimization can be performed numerically substituting \eqref{pvk} and \eqref{pvkbk} in \eqref{max1}. 
Let $I_{k,w_k}=I(B_k; V_k( q_k(w_k)))$  be the mutual information between $B_k$ and $V_k$ using $w_k$ bits and the quantization step $q_k(w_k)$ obtained in \eqref{eq:Sk}.
Considering the Gray mapping of DVB-C2 \cite{DVB-C2}, we can treat independently the real and the imaginary parts of the constellation points \cite{Hyun-dec05}. 
We map the bits $b_k$ on the imaginary axis when $k$ is odd. Similarly, we map the bits $b_k$ on the real axis when $k$ is even.
The symmetry introduced by the Gray mapping implies   $q_{2u-1} =  q_{2u}$, with $u=1, 2, \ldots, \frac{\log M}{2}$.

\begin{table}\centering
\caption{ Best $q_k$ which maximizes the MI for each $k$ and $w_k$ considering a 4,096-QAM and AWGN with C/N = 32.2 $\rm dB$ }
    \begin{tabular}{|cr||c|c|c|c|c|c|}
\hline
~ & ~ & \multicolumn{6}{|c|}{$k$}  \\
  ~ &     ~   & $1,2$  & $3,4$ & $5,6$ & $7,8$ & $9,10$ & $11,12$   \\ \hline \hline
\multirow{5}{*}{$w_k$} & 2 & 3.73&3.40&3.13&2.93&2.53&1.80 \\ \cline{2-8}
&3& 2.23&2.00&1.83&1.77&1.46&1.03\\ \cline{2-8} 
&4& 1.21&1.12&1.05&0.97&0.84&0.55\\ \cline{2-8}
&5& 0.75&0.66&0.61&0.52&0.47&0.28\\ \cline{2-8}
&6& 0.38&0.36&0.34&0.32&0.26&0.14\\ \hline
    \end{tabular}
\label{GMItable2}
\end{table}

\begin{table}\centering
\caption{ Best $q_k$ which maximizes the MI for each $k$ and $w_k$ considering a 4,096-QAM and Block Rayleigh Fading with C/N = 34 $\rm dB$}
    \begin{tabular}{|cr||c|c|c|c|c|c|}
\hline
~ & ~ & \multicolumn{6}{|c|}{$k$}  \\

     ~ &     ~   & $1,2$  & $3,4$ & $5,6$ & $7,8$ & $9,10$ & $11,12$   \\ \hline\hline
\multirow{5}{*}{$w_k$} &2 & 2.60&2.40&2.27&2.07&1.73&1.53 \\ \cline{2-8}
&3&1.29&1.26&1.20&1.11&1.00&0.86 \\ \cline{2-8}
&4&0.79&0.72&0.65&0.64&0.59&0.52\\ \cline{2-8}
&5&0.39&0.43&0.36&0.34&0.34&0.33\\ \cline{2-8}
&6&0.21&0.25&0.23&0.21&0.21&0.19\\ \hline
    \end{tabular}
\label{GMItable2f}
\end{table}
In Tables  \ref{GMItable2} and \ref{GMItable2f}, we report the results of \eqref{eq:Sk}, for  4,096-QAM and  both AWGN and Rayleigh fading channels.
Note that when $w_k=1$  we are considering a hard decision on the LLR, and the BGMI does not depend on $q_k$. 
\revi{rev3-12}{For non-Gray mappings, we cannot exploit the above symmetry, and $q_k$ must be optimized for even and odd values of $k$.}

\subsubsection*{Optimization of the bit lengths $w_k$}

After having optimized the quantization step $q_k$ for each $k$ and each $w_k$, our focus is to find the best $\bm{w}$ subject to (\ref{totbits}). Therefore, the optimization objective \eqref{probmax} becomes
\begin{equation}
\max_{  \bm{w} } \sum_{k=1}^{\log M} I(B_k; V_k)\,,  \, \text{s.t. (\ref{totbits}). } 
 \label{W}
\end{equation}

Our approach to solving \eqref{W} is to assign one bit at a time to the  $k^*$-th quantizer that yields the highest gain in terms of MI, so that   $k^*=  \argmax_{k} \{I_{k,w_k+1} - I_{k,w_k} \}$. 
Therefore, after having computed $q_k(w_k)$ and $I_{k,w_k}$ for each  $k=1, 2, \ldots, \log M$, and $w_k=1, 2, \ldots, W$, the optimization \eqref{probmax} is solved using Algorithm \ref{algo1}.

\begin{algorithm}
	\caption{Optimization of the bit lengths $w_k$.}
\label{algo1}
Initialize $\bm{w} = ( 0, \ldots , 0 ) $\;
 \For{u = 1:W }{
   $k^*=  \displaystyle \argmax_{k} \{I_{k,w_k+1} - I_{k,w_k} \}$\;
$  w_{k^*} = w_{k^*}+1 $.
}

\end{algorithm}
\revi{rev2-6}{We find it interesting that, if $I_{k,w_k}$ is an upper convex sequence of $w_k$, this greedy procedure is optimal, in the sense that it returns the same results as an exhaustive search.
 A proof of this statement is reported in the Appendix. Although we could not  prove the upper convexity of $I_{k,w_k}$ under general conditions, as remarked in Section \ref{numres}, we find that this property holds true in all the cases considered in this paper, for both AWGN and fading channels.}
\input{W_table.tex}
\input{W_table_Rayleigh.tex}
Tables \ref{GMItable3} and \ref{GMItable3f} show the results of this optimization for our study case with 4,096-QAM, respectively, for AWGN and Rayleigh fading channels.
Also in this case, the symmetry introduced by Gray mapping implies $w_{2u-1} =  w_{2u}$, with $u=1, 2, \ldots, \frac{\log M}{2}$. \revi{rev3-12-1}{For non-Gray mappings, we have different values of $w_k$ for each $k$.}

\section{LLR Compression}
\label{compsec}

The second part of this paper is based on the  observation that the quantized LLR levels are not uniformly distributed, therefore compression can  reduce the memory needed to store the LLRs.
Let $\bm{v}= ( v_1, v_2, \ldots, v_{\log M} )$ be the vector of the LLR quantized levels coming from the same received symbol.
In order to allow the symbol de-interleaver to move blocks of the same size, the compression procedure must represent each $\bm{v}$ with the same number of bits. 
Then, our task is to design a procedure that maps the $W$ bits representing $\bm{v}$  into  $\bar{N}$ bits.

With this purpose, we propose a lossy compression is performed in two steps: first we do a lossless entropy coding applied separately on each $v_k$; then, if the number of bits exceed $\bar{N}$ we perform a further LLR compression as described in the following.   

For the lossless compression, we apply Huffman coding \cite{Cover-book} at the output of each LLR quantizer.  Let $m_{k,v_k}$ be the length of the Huffman codeword that represents the level $v_k$. Then the  number of bits required to represent  $\bm{v}$ is 
\begin{equation}
N = \sum_{k=1}^{\log M} m_{k, v_k}\,.
\label{storage}
\end{equation} 
If  $N \leq \bar{N}$,   no further compression is needed. The vector $\bm{v}$ can be either stored as it is or potentially padded with zeros to make it of length $\bar{N}$.
Otherwise, we modify one or more quantizer outputs so that the new $N$ is smaller or equal to the target $\bar{N}$.
Clearly this operation will cause a performance loss that we can quantify in terms of GMI. Our aim is to minimize this loss while reaching the target length  $\bar{N}$. 

Let $\delta_{k,a,c}$ be the average GMI loss incurred when we replace the LLR quantized level  $v_k=a$ with another level, $\hat{v}_k=c$.
Note that, by replacing $v_k=a$ with $\hat{v}_k=c$, we obtain the new PMFs
\begin{equation} \label{newPMF}
p_{\hat{V}_k|B_k}(v_k|b_k) = \begin{cases}
p_{V_k|B_k}(v_k | b_k) & \hat{v}_k \neq a, c \,,\\
0 &  \hat{v}_k = a \,,\\
p_{V_k|B_k}(a| b_k)+ \\
p_{V_k|B_k}(c| b_k) &  \hat{v}_k = c\,,
\end{cases}
\end{equation}
\begin{equation} \label{newPMF2}
p_{\hat{V}_k}(v_k) = \begin{cases}
p_{V_k}(v_k) & \hat{v}_k \neq a, c\,, \\
0 &  \hat{v}_k = a \,,\\
p_{V_k}(a) + \\
p_{V_k}(c) &  \hat{v}_k = c\,.
\end{cases}
\end{equation}
Therefore, considering  \eqref{newPMF}, \eqref{newPMF2} and (\ref{max1}), the average GMI loss, $\delta_{k,a,b}$  is given by
\begin{equation}\label{delta_k}
\begin{split}
\delta_{k,a,b} =  &   p_{V_k|B_k}(a| b_k) \log \frac{p_{V_k|B_k}(a| b_k)}{p_{V_k}(a)} \\
  & + p_{V_k|B_k}(b| b_k) \log \frac{p_{V_k|B_k}(b| b_k)}{p_{V_k}(b)} \\ 
& -  \sum_{b_k=0}^1 \left(p_{V_k|B_k}(a| b_k)+ p_{V_k|B_k}(b| b_k) \right) \times \\
& \log \frac{p_{V_k|B_k}(a| b_k)+ p_{V_k|B_k}(b| b_k)}{p_{V_k}(a) + p_{V_k}(b)} \,.
\end{split}
\end{equation}
Note that $\delta_{k,a,b}$  is zero if $a=b$, otherwise is non-negative.

In order to reach the compression target $\bar{N}$, one or more  LLR quantized levels $v_k$ will be replaced with a new level $\hat{v}_k$, having a shorter representation. 
The problem is to find the vector $\hat{\bm{v}}= ( \hat{v}_1, \hat{v}_2, \ldots, \hat{v}_{\log M} )$  that minimizes the average GMI loss, while keeping $N \leq \bar{N}$. Mathematically we aim at solving 
\begin{subequations}
\begin{equation}
\min_{ \hat{v}_1, \ldots, \hat{v}_{\log M} } \sum_{k=1}^{\log M} \delta_{k,v_k,\hat{v}_k} \,,
\label{Lag12}
\end{equation}
s.t.
\begin{equation}
\sum_{k=1}^{\log M} m_{k, \hat{v}_k} \leq \bar{N} \,.
\label{totbits2}
\end{equation}
\label{probmax2}
\end{subequations}
This problem can then be seen as a multidimensional multiple-choice knapsack (MMCK) problem  \cite{Kellerer-book}.
Unfortunately, the MMCK problem is NP hard \cite{Kellerer-book}, thus we resort to the following greedy iterative approach.

\subsubsection*{Greedy LLR compression}
Starting from $\bm{v}$, at each iteration, the algorithm selects the substitution $v_k \rightarrow \hat{v}_k$  yielding the smallest average GMI loss, considering only the $\hat{v}_k$ such that $m_{k, \hat{v}_k}<m_{k, v_k}$. 
The length $N$ is decreased at least by 1 at each  iteration. We stop the procedure when $N \leq \bar{N}$. 
The iterative procedure works as described in Algorithm \ref{algo2}.

\begin{algorithm}
\caption{Greedy LLR compression}
\label{algo2}
Initialize $\hat{v}_1 = v_1, \hat{v}_2 = v_2, \ldots, \hat{v}_{\log M} = v_{\log M}$;

\While{(\ref{totbits2}) is not satisfied}{
\For{$k=1,\ldots,\log M$}{
\For{$\hat{v}_k^\prime=1,\ldots,L_k$}{
\If{$m_{k, \hat{v}_k^\prime} \geq m_{k,\hat{v}_k}$}{$\delta_{k, v_k, \hat{v}_k^\prime} = \infty$}
}}
$ \displaystyle (k^*, \hat{v}_{k^*}^{\prime *}) = \argmin_{\substack{k=1,\dots,\log M \\ \hat{v}_k^\prime=1,\dots,L_k}} \delta_{k, v_k, \hat{v}_k^\prime}$;

$\hat{v}_{k^*} = \hat{v}_{k^*}^{\prime *}$;
}

\end{algorithm}
%
We have two bounds on the number of iterations required for the convergence. On one hand, as at each iteration we set at least one value of $\delta_{k, v_k, \hat{v}_k}$ to $\infty$ we have
\begin{equation}
\mbox{\# iterations} \leq \sum_{k=1}^{\log M} L_k\,.
\end{equation}
On the other hand, as $N$ is decreased by at least one bit at each iteration, we have
\begin{equation}
\mbox{\# iterations} \leq \bar{N} - N\,,
\end{equation}
and usually this second condition provides the tightest bound.

\subsubsection*{Joint Optimization of $\, W$ and $\bar{N}$}
In the previous section we have provided a detailed design of both LLR quantization and compression.
Following the proposed scheme, the only two parameters we need to set in order to specify the quantization and compression procedure are $W$ and $\bar{N}$, which represent the number of bits at the output of the quantizer and of the compressor, respectively.
Only $\bar{N}$ determines the final size of the memory, but both of them have an impact on the performance. In fact, if $W$ is much higher than  $\bar{N}$,  we will have a higher GMI at the output of the quantizer, but the lossy compression will be aggressive and will introduce significant loss. 
We do not know an easy way to determine the best $W$ for a given $\bar{N}$. 
In the numerical results reported in Fig.s \ref{wcPlot} and \ref{wcPlotFading}, we  tested several values of $W$ for each $\bar{N}$ and chose the one that gives the best performance.


\section{Numerical Results}\label{numres}

We evaluate the performance of the proposed solutions on the DVB-C2 standard for cable television. This standard provides OFDM with 4,096 subcarriers, BICM with LDPC codes and symbol interleaver (a combination of frequency and time interleaving), which  fits the scheme of Fig. \ref{fig1}. In particular, the symbol interleaver is a row-column block interleaver, with a number of rows up to 16 OFDM blocks and with a number of columns up to 3,236 (corresponding to the maximum number of data symbols in a OFDM block). Various constellation sizes are provided, from 16-QAM up to 4,096-QAM with Gray mapping. Hence, in the worst case scenario, the interleaving block contains 51,776 data cells or 621,312 LLR values.
In the following, we will refer to the carrier to noise (C/N) ratio as the SNR on each subcarrier after OFDM demodulation.

\subsection{Quantization Performance}

\begin{figure}
\centering
\includegraphics{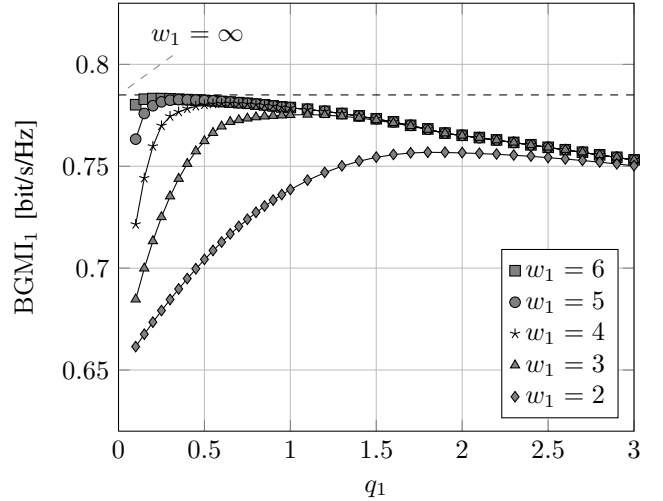}
\caption{BGMI of the quantized LLR of the MSB  for a 64-QAM constellation over AWGN channel with $C/N = 10$ dB, as a function of the quantization step $q_1$, for several values of $w_1$.  Lines show analytical results and  markers are obtained by Monte Carlo simulations.}
\label{GMIvsY1}
\end{figure}

\begin{figure}
\centering
\includegraphics{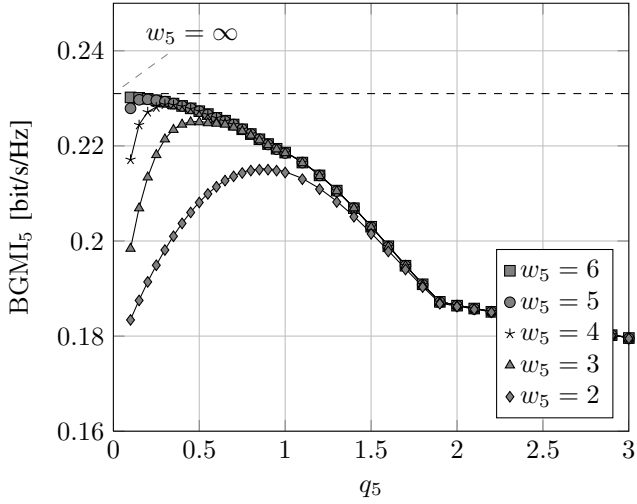}
\caption{BGMI of the quantized LLR of the LSB  of a 64-QAM constellation over AWGN channel with $C/N = 10$ dB, as a function of the quantization step $q_5$, for several values of $w_5$.  Lines show analytical results and markers are obtained by Monte Carlo simulations.}
\label{GMIvsY2}
\end{figure}

Fig.s \ref{GMIvsY1} and \ref{GMIvsY2} show the BGMI obtained from the quantized LLRs as a function of both  $q_k$ and $w_k$, for a C/N ratio of 10 dB, which represents the working point for 64-QAM. Results are reported for both the  least significant bit (LSB) and the most significant bit (MSB) along the real axis of 64-QAM symbols, i.e., for $k=1$ and $k=5$, respectively. Lines are obtained using the closed form expression of the PDF of the quantized LLRs, and markers show results obtained by Monte Carlo simulations. We see  perfect overlap between analytical and simulation results.

First, we note that for each value of $w_k$ we have only one optimum value of the quantization step $q_k$, which maximizes the BGMI. 
Then, we observe that both the maximum BGMI and the corresponding values of $q_k$ are different for the LSB and MSB. The same holds also for the other data bits (results are not reported here), with a behaviour similar to that of Fig.s \ref{GMIvsY1} and \ref{GMIvsY2}.  This justifies the use of different quantization steps for each bit of the constellation.
We note also that, as the number of bits $w_k$ increases, the maximum BGMI gets closer to the BGMI obtained with unquantized LLR, and the gain obtained using $w_k+1$ bit instead of $w_k$ gets smaller.
Also, for large quantization steps, the number of bits $w_k$ does not affect the BGMI performance, because adding bits provides quantization intervals for large values of LLR that do not contribute significantly to the BGMI.
%
\revi{rev3-25}{From Fig.s \ref{GMIvsY1} and \ref{GMIvsY2},  we also observe that it is important to characterize the LLRs close to zero: indeed, the distribution of LLR values around zero is also dominating the BER performance of uncoded systems \cite{Benjillali-07, Alvarado-09}.}

\begin{figure}
\centering
\includegraphics{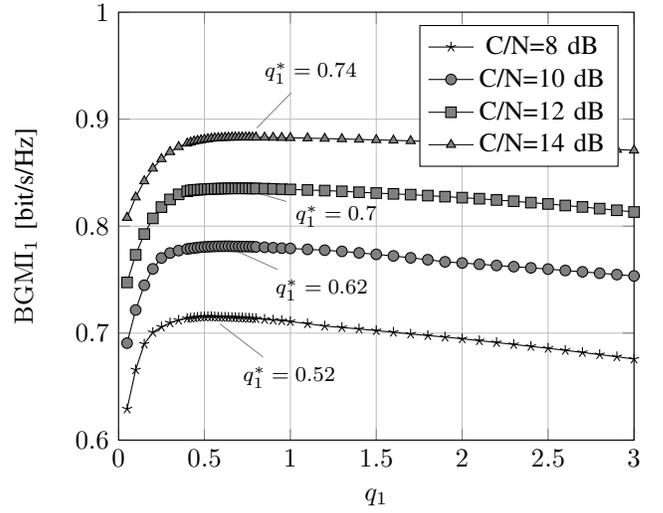}
\caption{BGMI of the quantized LLR of the MBS of a 64-QAM constellation using $w_1$=4 bits, as a function of the quantization step $q_1$ for several C/N values over AWGN channel.}
\label{GMIvsCN}
\end{figure}

\revi{rev2-7}{Fig. \ref{GMIvsCN} plots  BGMI values for the MSB of 64 QAM quantized with 4 bits, as a function of both $q_1$ and C/N. We note that the higher the C/N is, the larger the quantization step $q_1^*$ that maximizes the BGMI is. In fact, as the PDFs of the LLRs shift towards higher absolute values as the C/N increases, for larger C/N, it pays off to enlarge the quantization range at the expense of a coarser quantization near zero.}

We then consider larger constellations, in particular the 4,096-QAM constellation used in DVB-C2, which represents the worst-case scenario for the symbol interleaver memory size.  
The following results were obtained by considering C/N = 32.2 dB for AWGN and C/N = 34 dB for Rayleigh fading, because, according to \cite[Table 20, p. 128]{DVB-C2}, it represents the lowest working points for the 4,096-QAM. 
In Tables  \ref{GMItable2} and \ref{GMItable2f}, we report the optimal $q_k$ solving \eqref{eq:Sk}, for $w_k = 2, 3, \ldots ,  6$, and for each LLR position of the 4,096-QAM constellation, $k$, respectively in AWGN, and Rayleigh fading conditions.

\begin{figure}\centering
\includegraphics{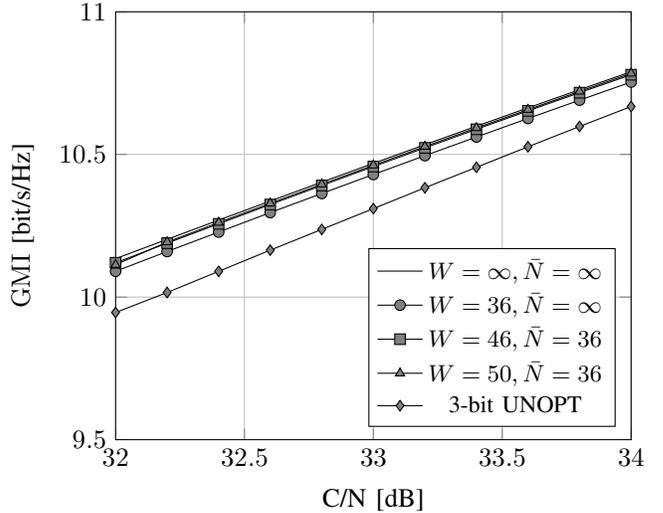}
\caption{GMI as a function of the C/N using $\bar{N} = 34$ and different values of $W$, considering 4,096-QAM over AWGN channel. 3-bit UNOPT: unoptimized system.}\label{GMIvsSNR}
\end{figure} 
\revi{rev2-8}{As the LLR statistics depend on the C/N, it is possible to adapt quantization according to (\ref{probmax}). 
However, to avoid re-computing (\ref{probmax}), we propose to use tables for $q_k$ and $w_k$,  computed considering the lowest working point.
Fig. \ref{GMIvsSNR} shows that GMI increases as the  C/N increases, even if $q_k$ and $w_k$ are computed considering the lowest working point (which in this case is 32.2 dB) rather than the actual C/N. Therefore when C/N is higher than the lowest working point, the required performance is certainly reached at any rate.}

\begin{figure}\centering
\includegraphics{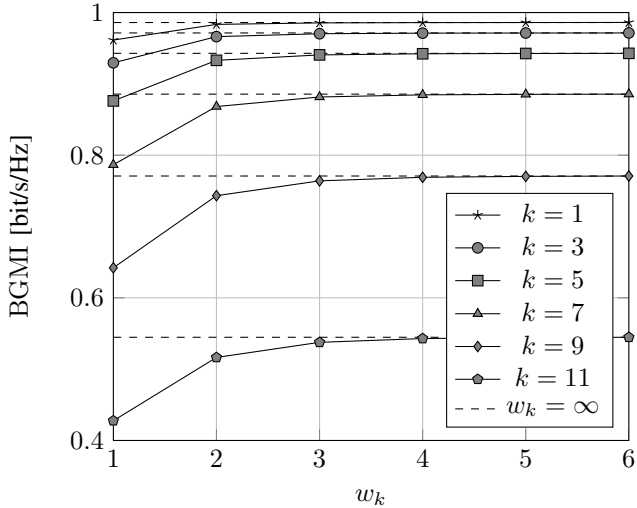}
 \caption{BGMI of the quantized LLR for different values $w_k$ and using optimal quantization step $q_k$, considering 4,096-QAM over AWGN channel with $C/N = 32.2$ dB.}
\label{GMIVsI}
\end{figure} 
In Fig. \ref{GMIVsI}  the maximized BGMI for each bit and for each value of $w_k$ are shown.
Again, we observe that the BGMI is significantly different for each bit of the constellation and also that the gain achieved by adding quantization levels is different for each bit. For example, going from $w_k=1$ to $w_k=6$ for the MSB provides an increase of BGMI of about 0.025 bit/s/Hz, while for the LSB we have a BGMI gain of 0.12 bit/s/Hz. \revi{rev3-38}{Therefore, for a given number of total available bits $W$, the maximum GMI is obtained by assigning a different number of bits $w_k$ to each constellation bit $k$, as discussed in Section III.b.} Furthermore, as also noted in Fig.s \ref{GMIvsY1} and \ref{GMIvsY2}, the BGMI  is an upper convex sequence of $w_k$, therefore the proposed algorithm for solving (\ref{W}) returns the same results of an exhaustive search.
Lastly, in Tables  \ref{GMItable3} and  \ref{GMItable3f}, we report the results of the $\bm{w}$ optimization, showing the optimal distribution of bits $w_k$ by solving (\ref{probmax}), for both  AWGN and Rayleigh fading channels. 
As expected, we observe that a finer quantization (i.e., higher $w_k$) of the LLR associated with LSB bits, which are less protected by the Gray mapping, pays off.

\subsection{Quantization and Compression Performance}

\begin{figure}\centering
\includegraphics{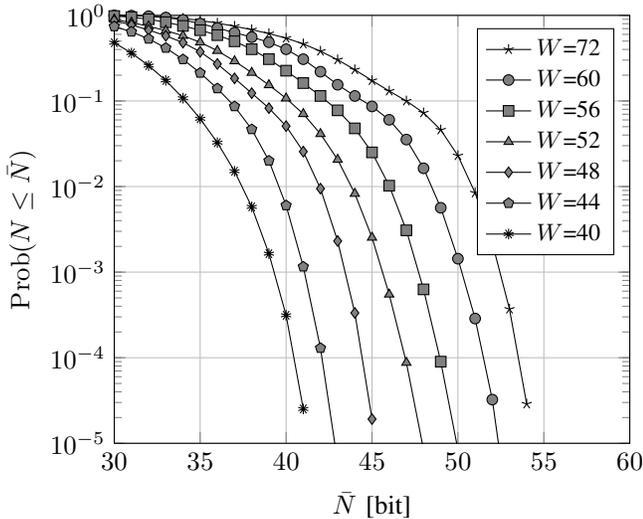}
\caption{CCDF of the word encoded length $N$.}\label{HuffmanCDF}
\end{figure} 

We now evaluate the effect of the LLR quantization and compression in terms of SNR gap, i.e., the amount of additional transmit power (or noise power reduction) required when quantization is used in order to achieve the same GMI of a receiver operating without quantization.

Fig. \ref{HuffmanCDF} shows the complementary cumulative distribution function (CCDF) of the encoded word length $N$ for different values of $W$. We observe that the Huffman coding provides a significant reduction of the number of required bits to describe the quantized LLR. For example, for $W=72$, in 90\% of the realizations $N \leq 47$, with a compression of about 50\%. For $W=60$ the probability of having $N > 50$ is less than 0.001.

Hereafter, we show the GMI performance of the optimized quantization as a function of the C/N. First we note that the optimal quantization step depends on both the C/N itself and the channel conditions.
Usually, the performance of DVB-C2 is assessed by providing the minimum C/N at which a given BER is achieved. In terms of GMI, we can compare different solutions by considering the minimum C/N at which a given GMI is achieved.
In practice, we can optimize both $\bm{q}$ and $\bm{w}$ considering the lowest C/N at which a target GMI is achieved as higher  C/N values will not decrease the GMI.

Fig. \ref{GMIvsSNR} shows the GMI as a function of the C/N for various values of $W$, but with the same value of $\bar{N}=36$ bits, hence for the same interleaver memory size.
We also show the performance of the unoptimized system (3-bit UNOPT)  where the same 3-bit quantizer is used for all data bits of the constellation. Also for UNOPT, the total number of bit for constellation point is 36.
We observe that by using the iterative compressing procedure of Section \ref{compsec}, we do not incur  any significant loss in terms of GMI. In our example, the outputs of the optimized quantization using  $W=50$, and  $W=46$ bits, respectively, have been compressed to  $\bar{N}=36$ bits, thus outperforming the case of a sheer quantization using $W=36$ bits. 
\begin{figure}\centering
\includegraphics{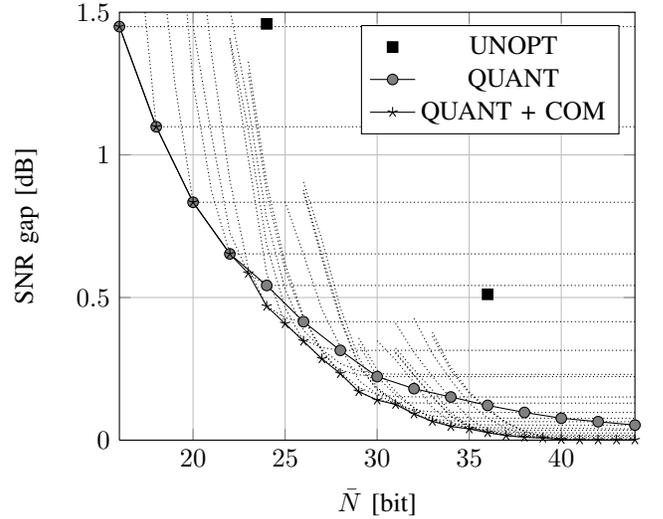}
\caption{SNR gap for quantized and quantized and compressed LLR as a function of $\bar{N}$, for 4,096-QAM at $C/N = 32.2$ dB over AWGN channel. UNOPT: unoptimized system; QUANT: system with quantized LLR; QUANT + COM: system with quantized and compressed LLR.}\label{wcPlot}
\end{figure} 

\begin{figure}\centering
\includegraphics{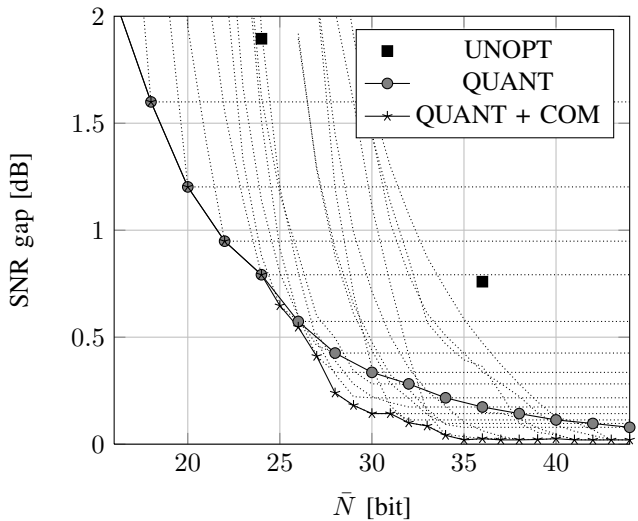}
\caption{SNR gap for quantized and quantized and compressed LLR as a function of $\bar{N}$, for 4,096-QAM at $C/N = 34$ dB, in block Rayleigh fading channel. UNOPT: unoptimized system; QUANT: system with quantized LLR; QUANT + COM: system with quantized and compressed LLR.}\label{wcPlotFading}
\end{figure} 

Another comparison between the system with quantization (QUANT) and the system with quantization and compression (QUANT + COMP) is provided in Fig.  \ref{wcPlot}, where the SNR gap is reported as a function of the total number of compressed bits $\bar{N}$, thus as a function of the required memory.
The dotted lines represent the SNR gap in case of QUANT+COMP for different values of $W$. 
In other words, each line represents the performance of the optimized quantizers using $W$ bits, where the output is then compressed from $W$ to $\bar{N}$  bits.
We note that for any of these curves the SNR gap decreases as $\bar{N}$ increases, because the loss due to compression is reduced, until $\bar{N} = W$, when compression has no effect and the SNR gap flattens. 
The line with star markers shows the minimum SNR gap achievable by QUANT+COMP approach. This result is obtained by choosing the $W$ that  reaches the minimum SNR gap, for each values of  $\bar{N}$. 
The QUANT case performance is shown with gray circle markers, in this case, as there is no compression we consider $\bar{N} = W$.
Finally, square black markers show the performance on an unoptimized system (UNOPT), where the same quantizer is used for LLRs of all data bits. 
In this case, as $w_k$ is constant for all $k$, $W$ can be only a multiple of $\log M$.

We observe that the optimization of both quantization and compression provides a significant reduction of the SNR gap with respect to a traditional unoptimized system. 
As shown in Fig. \ref{wcPlot}, the optimized quantization, QUANT, outperforms the unoptimized quantization, UNOPT,  with a gain of 0.8 dB and 0.4 dB, for $\bar{N}=24$, and $\bar{N}=36$ respectively.
Interestingly, the use of compression yields an advantage only for large values of $\bar{N}$. For example, if we target a SNR gap of 0.1 dB we need $\bar{N} = 32$ bit with QUANT+COMP, whereas we need $\bar{N} = 38$ bit with QUANT.

Note that the use of compression yields advantages only if the loss target is small enough.  For example, if we target a SNR gap larger than 0.7 dB, the QUANT + COMP approach does not bring any gain with respect to the QUANT approach. In other words, it is not efficient to compress LLRs that are already quantized optimally by using a limited number of bits. 

Fig. \ref{wcPlotFading} shows the comparison between QUANT, QUANT + COMP, and UNOPT in the case of a block Rayleigh fading channel. Here, the SNR gap is computed at C/N = 34 dB (different from AWGN), because the C/N working point in this case is higher. Also in this case if the target SNR gap is 0.2 dB, we need $\bar{N} = 29$ bit with QUANT+COMP, and $\bar{N} = 34$ bit with QUANT.
The performance gap between the optimized and the unoptimized quantization is even more significant in the case of a block Rayleigh fading channel. In fact, QUANT shows a SNR gain of 1.1 dB and 0.6 dB in the case of $\bar{N}=24$ and $\bar{N}=36$, respectively.

\subsection{BER Comparison}

\begin{figure}\centering
\includegraphics{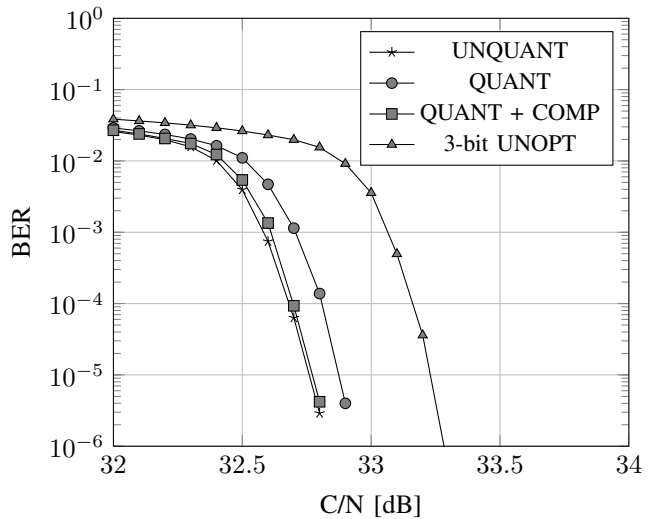}
\caption{BER vs C/N. UNQUANT: unquantized QAM; QUANT: system with quantized LLR; QUANT + COM: system with quantized and compressed LLR; UNOPT: unoptimized quantization; All the quantized system use a total of $\bar{N}=36$ bits.  }
\label{BER}
\end{figure}

In order to understand the effect of quantization on a system that uses specific error correcting codes, we obtained the BER of a DVB-C2 system by using LDPC codes with block length 64K, code rate 5/6, and 4,096 QAM constellation \cite{DVB-C2}.
Fig. \ref{BER} shows the comparison in terms of BER between QUANT and QUANT + COMP.
Both use 36 bits in total. For comparison, we also included the case of unquantized QAM (UNQUANT), i.e., $W = \infty$, and the  unoptimized case (3-bit UNOPT)  where the same 3-bit quantizer is used for all data bits of the constellation, thus it uses in total 36 bits. 

We observe that the losses due to quantization and compression agree with the SNR gap computed thought the GMI and illustrated in Fig. \ref{wcPlot}. In particular, the UNOPT system has a 0.5 dB loss with respect the UNQUANT case. This loss decreases to about 0.15 dB using QUANT, and becomes negligible when using QUANT + COMP.

\subsection{Memory Comparison}
\label{memcomp}

\begin{table*} \caption{Memory Comparison.}\label{t:MEM}
\centering
    \begin{tabular}{|l|l|c|c|c|c|c|c|c|c|}
        \hline
        Loss  & Receiver  & \multirow{2}{*}{$B_S$} & \multirow{2}{*}{$B_H$} & \multirow{2}{*}{$W$} & \multirow{2}{*}{$\bar{N}$} & $\Sigma(M_{\mathrm{SD}})$  & $\Sigma(M_{\mathrm{BD}})$  & $\Sigma(M_{\mathrm{Tot}})$   &  Saved   \\
       Target & Scheme & & & & & [Mbit] & [Mbit] & [Mbit] &Memory \\ \hline
     \multirow{3}{*}{0.1 dB}  &     CONV     &  15    & 14     & 60   & -   &2.27 &0.32 &2.60  & -                       \\ 
           & QUANT    & -     & -  & 38  & -        & 1.97      & 0.2      & 2.17        & 16.5               \% \\
             & QUANT + COMP     & -     & -     &  42   &    32      & 1.66&0.22&1.88    & 27.6     \%           \\  
           \hline
          \multirow{3}{*}{0.2 dB}&  CONV     & 14    & 13     &  60  & -  & 2.12& 0.32& 2.44     & -                    \\ 
    & QUANT    & -     & -  & 32  & -        & 1.66      & 0.17      & 1.83        & 25.2               \% \\
        & QUANT + COMP              & -   & -  & 36  & 29         & 1.50      & 0.19      & 1.69        & 30.6               \% \\ 
        \hline
    \end{tabular}
\end{table*}

We now compare  the conventional scheme (CONV) illustrated in Fig. \ref{fig2} (a),  and  the proposed scheme QUANT + COMP, illustrated in Fig. \ref{fig2}b, in terms of required memory. We assume that all de-interleavers are designed such that they can be written and read simultaneously.

In CONV, for each data cell, the received complex symbol, $r_i$ and the channel estimate, $h_i$, have to be stored in memory $M_{\mathrm{SD}}^{a}$. 
In order to save memory, the receiver can compensate the phase rotation due to the channel  after its estimation and then simply store  the magnitude of the channel estimates.
Therefore the size of memory $M_{\mathrm{SD}}^{a}$ is
\begin{equation}
\Sigma(M_{\mathrm{SD}}^{a}) = N_S (2 B_S + B_H)\,,
\end{equation}
where $N_S$ is the number of data cells to be interleaved, $B_S$  is the number of bits per axis to represent $r_i$, and  $B_H$ is the number of bits to represent  $h_i$. 
Whereas, in the proposed scheme, the compressed LLRs associated with one data cell occupies at most $\bar{N}$ bits, then the size of memory $M_{\mathrm{SD}}^{b}$ is
\begin{equation}
\Sigma(M_{\mathrm{SD}}^{b}) = N_S \bar{N}\,.
\end{equation}
The memory size for the bit interleaver $M_{\mathrm{BD}}$ in both schemes is
\begin{equation}
\Sigma(M_{\mathrm{BD}}) = \frac{N_B W}{\log M} \,,
\end{equation}
where $N_B$ is the depth of the bit interleaver.
Note that here the compressing  procedure is not applicable because the LLRs are moved one by one by the bit interleaver, therefore each LLR $\hat{v}_{k}$ will be represented by $w_k$ uncompressed bits. 
For DVB-C2, the maximum value of $N_B$ is 64,800, and for the symbol interleaver $N_S$ is at most  51,776. Therefore in DVB-C2 the size of $M_{\mathrm{SD}}$ overrides that of $M_{\mathrm{BD}}$.
In all the following assessments, we will consider the worst case, 4,096-QAM, which maximizes the size of $M_{\mathrm{SD}}^{b}$.
DVB-C2 performance assessments show that in order to to have a SNR gap smaller than $0.1$ dB, we have to use at least $B_S = 15$ and $B_H = 14$ bit to represent $r_i$ and $h_i$, respectively, and $w_k=5$  bit for each LLR. Thus the total required memory size, $\Sigma(M_{\mathrm{Tot}}) = \Sigma(M_{\mathrm{SD}})+\Sigma(M_{\mathrm{BD}})$,  is around 2.6  Mbit.
On the contrary, in the proposed  scheme we are able to reach the same target using the compressing procedure with parameters $W=42$, and $\bar{N}=32$. The total memory size becomes 1.88 Mbit, thus providing 27.6\% of memory saving.  Note that in the proposed scheme we can represent  $r_i$ and $h_i$ by using as much precision as needed to have a negligible loss. The values of $B_S$ and  $B_H$ will have no effect on the interleaver memory size.
If the  target on the SNR gap is more relaxed, for instance $0.2$ dB,  the saved memory becomes even larger. In fact, in CONV, to obtain a SNR gap smaller than $0.2$ dB, we need $B_S = 14$, $B_H = 13$,  and $w_k=5$,  thus the total required memory is around 2.44  Mbit.
Whereas, in QUANT + COMP, the target is achieved using $W=36$, and $\bar{N}=29$ compressed bit for data cell, and then requiring about 1.69 Mbit, therefore achieving a memory reduction of more than 30\%. It is interesting to note that, also in case of no compression (i.e. QUANT), the total memory size is reduced by more than 25\% with respect to the conventional receiver.
The required memory size and the potential memory saving are summarized in Table \ref{t:MEM}.

\section{Conclusions}

In this paper we have proposed and analyzed a new technique for the quantization and compression of LLR in a communication system that uses long interleavers.
The proposed quantization yields a memory size reduction of at least 16\% with negligible increase of the complexity. Quantization and compression reduce the memory size by up to  30\%.

\appendix

In the following, we report the proof of the optimality of the greedy procedure, in case of upper convexity  of $I_{k,w_k}$. 
\begin{IEEEproof}
Let $\delta_{i,j} =I_{i,j} -  I_{i,j-1}$ be the elements 
of a matrix $\Delta=\{\delta_{i,j}\}$ having dimension  $ \log M \times W $.
Since,  $0 \leq \delta_{i,j} \leq \delta_{i,j-1} \; \forall i$  each row of $\Delta$ is a non-increasingly sorted vector.
We can rewrite the optimization  \eqref{W} as follows,
\begin{equation} \label{Wproof}
\max_{\{ \bm{w}\}}  \sum_{i=1}^{\log M} \sum_{j=1}^{w_i}  \delta_{i,j}  \, \text{s.t. (\ref{totbits}). } 
\end{equation}
Clearly the optimization objective is maximized when the largest $W$ elements  of  matrix $\Delta$ are summed.
Let $\delta_{[\ell]}$ be the $\ell$-th largest element of $\Delta$, then we can write the maximized 
optimization objective as
\begin{equation} \label{Wproof2}
 \sum_{\ell=1}^{W} \delta_{[\ell]}  \, .
\end{equation}
Assuming that $\tilde{\bm{w}}=( \tilde{w}_1, \ldots , \tilde{w}_{\log M} )$ is the vector that maximizes \eqref{W} using $\tilde{W}$ bits, we can write the  $(\tilde{W}+1)$-th largest element of $\Delta$ as
\begin{equation}
\delta_{[\tilde{W}+1]}  = \max_{i} \left\{ \max_{j>\tilde{w}_i}  \delta_{i,j} \right\}  \, .
\end{equation}
Since  $\delta_{i,j} \leq \delta_{i,j-1} $, it becomes 
\begin{equation}
\delta_{[\tilde{W}+1]}  =  \max_{i} \delta_{i,\tilde{w}_i +1}  \, .
\end{equation}
That is precisely the rule used in our procedure.
Therefore the proposed procedure will distribute the remaining $W$ bits in an optimal way, i.e., returning the same result of an exhaustive search.
\end{IEEEproof}

\end{document}

%% file: W_table.tex
\begin{table*}\footnotesize 
 \caption{ Optimal bit distribution sets for 4096-QAM, considering AWGN and C/N = 32.2 $\rm dB$}
\centering
  \begin{tabular}{|c|c|c|c|c|c|c|c|c|c|c|c|c|c|c|c|c|c|c|c|c|}
           \hline
\multirow{6}{*}{
 $
 \begin{matrix}
 w_1, w_2 \\
 w_3, w_4 \\
 w_5, w_6 \\
 w_7, w_8 \\
 w_9, w_{10} \\
 w_{11}, w_{12} 
 \end{matrix} \! \!$
}
&1&1&1&1&1&1&1&1&2&2&2&2&2&2&2&3&3&3&3&3 \\
&1&1&1&1&1&2&2&2&2&2&2&2&3&3&3&3&3&3&4&4\\
&1&1&1&1&2&2&2&2&2&2&3&3&3&3&3&3&4&4&4&4\\
&1&1&1&2&2&2&2&2&2&3&3&3&3&3&4&4&4&4&4&4\\
&1&2&2&2&2&2&3&3&3&3&3&4&4&4&4&4&4&5&5&5\\
&1&1&2&2&2&2&2&3&3&3&3&3&3&4&4&4&4&4&4&5\\ \hline
 $W$ & 12&14&16&18&20&22&24&26&28&30&32&34&36&38&40&42& 44&46&48&50 \\ \hline
      \end{tabular}
 \label{GMItable3}
\end{table*}

%% file: W_table_Rayleigh.tex
\begin{table*}\footnotesize 
 \caption{ Optimal bit distribution sets for 4096-QAM, considering Rayleigh Fading and C/N = 34 $\rm dB$}
\centering
  \begin{tabular}{|c|c|c|c|c|c|c|c|c|c|c|c|c|c|c|c|c|c|c|c|c|}
           \hline
\multirow{6}{*}{
 $
 \begin{matrix}
 w_1, w_2 \\
 w_3, w_4 \\
 w_5, w_6 \\
 w_7, w_8 \\
 w_9, w_{10} \\
 w_{11}, w_{12} 
 \end{matrix} \!  \!$
}
&1&1&1&1&1&1&1&2&2&2&2&2&2&2&2&2&3&3&3&3\\
&1&1&1&1&1&2&2&2&2&2&2&2&2&3&3&3&3&3&3&4\\
&1&1&1&1&2&2&2&2&2&2&3&3&3&3&3&4&4&4&4&4\\
&1&1&1&2&2&2&2&2&2&3&3&3&3&3&4&4&4&4&4&4\\
&1&2&2&2&2&2&2&2&3&3&3&3&4&4&4&4&4&5&5&5\\
&1&1&2&2&2&2&3&3&3&3&3&4&4&4&4&4&4&4&5&5\\\hline
 $W$ &12&14&16&18&20&22&24&26&28&30&32&34&36&38&40&42 &44&46&48&50\\ \hline 
     \end{tabular}
 \label{GMItable3f}
\end{table*}